\documentclass[prl,twocolumn,notitlepage,amsfonts,amssymb,nofootinbib,superscriptaddress]{revtex4-1}
\usepackage{graphicx}
\usepackage{color}
\usepackage{amsmath, amsthm, amssymb}
\usepackage{float}
\usepackage{natbib}

\newcounter{addendumfig}
\begin{document}
\bibliographystyle{apsrev}
\title{THz response and colossal Kerr rotation from the surface states of the topological insulator Bi$_2$Se$_3$}



\author{R. Vald\'es Aguilar}
\email{rvaldes@pha.jhu.edu}
\affiliation{The Institute for Quantum Matter, Department of Physics and Astronomy, The Johns Hopkins University, Baltimore, MD 21218 USA.}
\author{A.V. Stier}
\affiliation{Department of Physics, University at Buffalo. State University of New York. Buffalo, NY 14260}
\author{W. Liu}
\affiliation{The Institute for Quantum Matter, Department of Physics and Astronomy, The Johns Hopkins University, Baltimore, MD 21218 USA.}
\author{L.S. Bilbro}
\affiliation{The Institute for Quantum Matter, Department of Physics and Astronomy, The Johns Hopkins University, Baltimore, MD 21218 USA.}
\author{D.K. George}
\affiliation{Department of Physics, University at Buffalo. State University of New York. Buffalo, NY 14260}
\author{N. Bansal}
\affiliation{Department of Physics and Astronomy, Rutgers the State University of New Jersey. Piscataway, NJ 08854}
\author{L. Wu}
\affiliation{The Institute for Quantum Matter, Department of Physics and Astronomy, The Johns Hopkins University, Baltimore, MD 21218 USA.}
\author{J. Cerne}
\affiliation{Department of Physics, University at Buffalo. State University of New York. Buffalo, NY 14260}
\author{A.G. Markelz}
\affiliation{Department of Physics, University at Buffalo. State University of New York. Buffalo, NY 14260}
\author{S. Oh}
\affiliation{Department of Physics and Astronomy, Rutgers the State University of New Jersey. Piscataway, NJ 08854}
 \author{N.P. Armitage}
 \email{npa@pha.jhu.edu}
 \affiliation{The Institute for Quantum Matter, Department of Physics and Astronomy, The Johns Hopkins University, Baltimore, MD 21218 USA.}
 
\begin{abstract}
We report the THz response of thin films of the topological insulator Bi$_2$Se$_3$.  At low frequencies, transport is essentially thickness independent showing the dominant contribution of the surface electrons. Despite their extended exposure to ambient conditions, these surfaces exhibit robust properties including narrow, almost thickness-independent Drude peaks, and an unprecedentedly large polarization rotation of linearly polarized light reflected in an applied magnetic field. This Kerr rotation can be as large as 65$^\circ$ and can be explained by a cyclotron resonance effect of the surface states.
\end{abstract} 
 
\maketitle

Ordered states of matter are typically categorized by their broken symmetries. With the ordering of spins in a ferromagnet or the freezing of a liquid into a solid, the loss of symmetry distinguishes the ordered state from the disordered one.  In contrast, topological states are distinguished by specific \textit{topological} properties that are encoded in their quantum mechanical wavefunctions \cite{TKNN1982}.  Frequently, a consequence of these properties is that there are robust ``topologically protected" states on the sample's boundaries.   The edge states of the quantum Hall effect (QHE) are the classic example \cite{Klitzing80a}.  In the last few years, it was realized that another class of such topological matter may exist in 3D band insulators with large spin-orbit interaction \cite{Bernevig05a,Fu07a,Moore07a,Roy09a}.  These so-called topological insulators are predicted to host robust surface states, which exhibit a number of interesting properties including spin helicity, immunity to back-scattering, and weak \textit{anti}--localization.  There are predictions of a number of unusual phenomena associated with these surface states, including a proximity-effect-induced exotic superconducting state with Majorana fermions bound to a vortex \cite{Fu08a,Akhmerov09} and an \textit{axion} electromagnetic response \cite{Qi08b,Essin09a}, and proposals for applications, such as their use in terahertz (THz) devices \cite{ZhangApp}.  

Most of the signatures of topological behavior in these materials thus far have come from surface probes such as angle resolved photoemission (ARPES) and scanning tunneling spectroscopy \cite{Hsieh08a,Xia09a,Hsieh09b,Chen09a,Roushan09a,Alpichshev:2010a}. These experiments have revealed that the surface states indeed show signatures of the predicted topological properties, such as a Dirac-like dispersion, chiral spin textures, and the absence of backscattering.  Direct observation of the topological behavior in transport has been hampered by the lack of a true bulk insulating state.  Only recently have transport experiments started to distinguish the surface contribution from the bulk \cite{Qu10a, Butch10a,Xiong:2011fk,Bansal11a}. 

As opposed to the case of the quantum Hall effect, in topological insulators, the quantization of the off-diagonal conductivity is not a requirement for the existence of the topological state.  This, along with the problem of bulk conduction, has made finding a unique signature of this state difficult. It has been proposed that topological insulators may be characterized by their electrodynamic properties \cite{Qi08b} due to the existence of an \textit{axionic} term in the action $\Delta \mathcal  L = \alpha \theta \int dxdt\mathbf{E}\cdot\mathbf{B}$, where $\alpha$ is the fine structure constant.  $\theta$ is a modulo $2\pi$ number that distinguishes between topologically non-trivial ($\theta$=$\pi$) and trivial insulators ($\theta$=0). There have been several proposals of how to measure this $\theta$ parameter using polarized electromagnetic radiation at low frequencies \cite{Qi08b,Tse10a,Tse10b,Maciejko10a,Lan11a,Tkachov10a}.  All these proposals predict that linearly polarized THz range electromagnetic waves reflected from (transmitted through) a sufficiently low disorder topological insulator in a magnetic field, will undergo a Kerr (Faraday) rotation of the polarization plane, the magnitude of which depends on the experimental configuration, but which is set by the fine structure constant $\alpha$ itself.  Thus, such experiments can directly probe the topological nature of the surface states.

\begin{figure}[t]
\includegraphics[width=1\columnwidth]{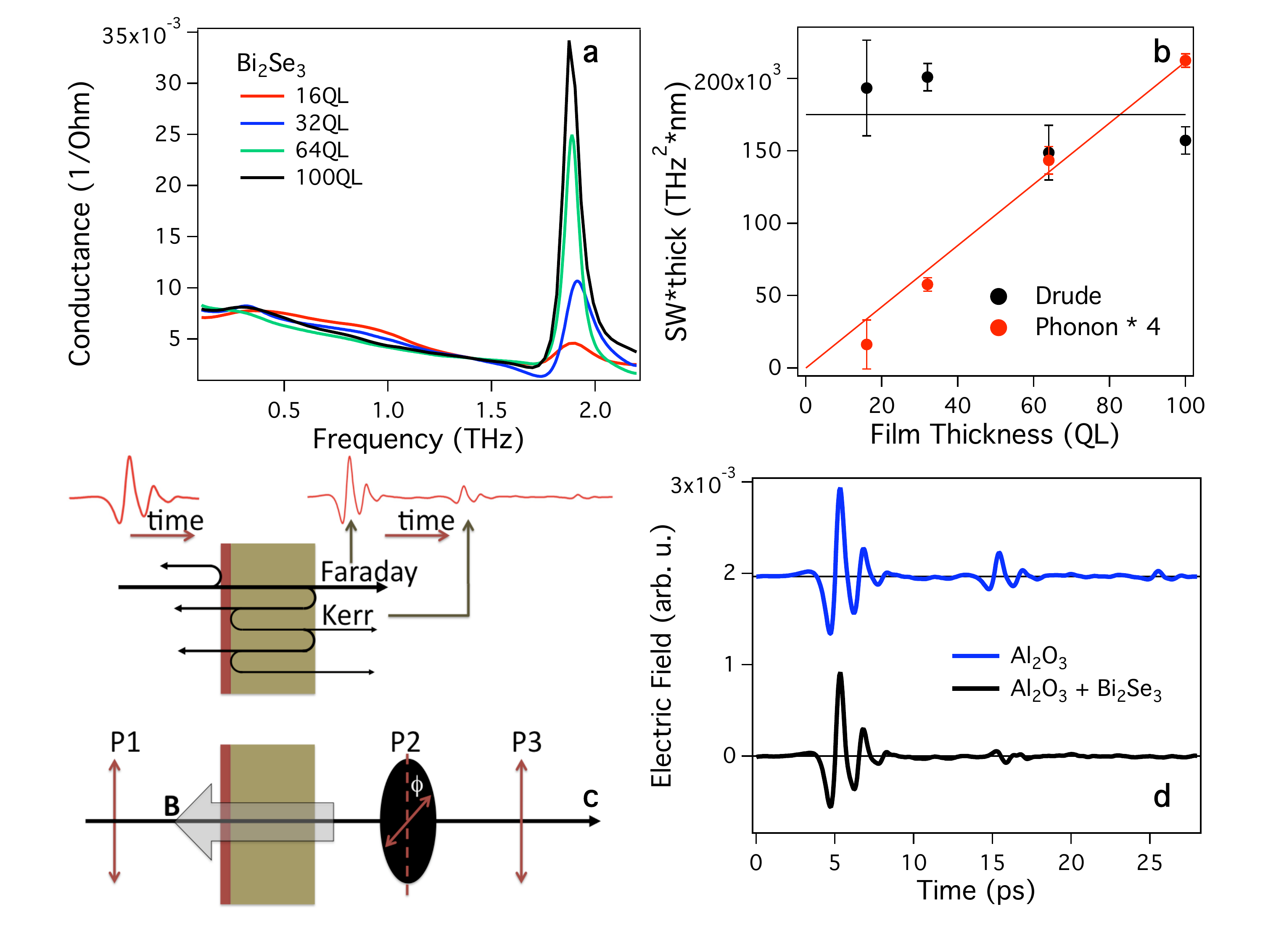}
\caption{\textbf{a}). Real part of the conductance of three films of different thicknesses (16, 32 and 100 QL, 1 QL=9.4\AA). \textbf{b}). Spectral weight of the Drude and the phonon contributions obtained from the fit to the conductance (symbols). Lines are guides to the eye. \textbf{c}). Schematic of experiment, where multiple echoes may be separated in time (top). The bottom figure shows the polarizer arrangement for experiments in a magnetic field, the first and last polarizers can be set either parallel or perpendicular to each other, and the middle one can be rotated.   \textbf{d}). Time domain trace of the transmission through a bare Al$_2$O$_3$ (top, displaced vertically for clarity), and a thin film sample (bottom) at 2 K.}
\label{Fig1}
\end{figure}

In this Letter, we report the THz response of the topological surface states (TSS) in thin films of the topological insulator Bi$_2$Se$_3$.  We measured thin films of several thicknesses grown on sapphire (Al$_2$O$_3$) substrates by molecular beam epitaxy.  Due to their small thickness, low bulk carrier density and high mobilities, these films have been recently shown to exhibit thickness-independent DC transport \cite{Bansal11a}. We find clear signatures of the protected surface states in the 2D behavior of the THz conductivity obtained using time domain terahertz spectroscopy (TDTS). In addition we find a colossal Kerr rotation almost entirely due to the TSS. The Kerr rotation measurement allows us to extract the effective mass of the 2D Dirac electrons. 

In Fig. \ref{Fig1}a we show typical data of the real part of the longitudinal conductance (G$_{xx}$ = $\sigma_{xx} t$) of a number of samples of different thicknesses ($t$) at 6 K and at zero magnetic field.  The data show a clear signature of free electron behavior with a reasonably narrow ($\approx$ 1.2 THz wide) Drude peak centered at zero frequency.   In addition, we observe a contribution from an optical phonon close to 2 THz. The data are qualitatively similar to ones reported by others in single crystals \cite{Laforge10a,Sushkov10a}. We fit these data with a model consisting of two identical Drude terms (one for each surface), and a Drude-Lorentz term for the bulk phonon.  These fits allow an essentially perfect parameterization of the data (see ref.\cite{SuppMat}) and show that the surface Drude term has an almost thickness-independent integrated spectral weight (Fig. \ref{Fig1}b). The phonon spectral weight shows linear dependence with thickness, typical of bulk response. In this fashion we conclude that the surface transport dominates the free electron response at these thicknesses.  A similar conclusion has been reached through DC studies of these films \cite{Bansal11a}.  Using the Fermi velocity from ARPES \cite{Hsieh09b}, our observed scattering rate gives a  mean free path of approximately 4.5 microns. The small scattering rate and long mean free path are remarkable considering that these films have no capping layers and the `active area' has been directly exposed to atmosphere for several days. 

\begin{figure}[t]
\includegraphics[width=.85\columnwidth]{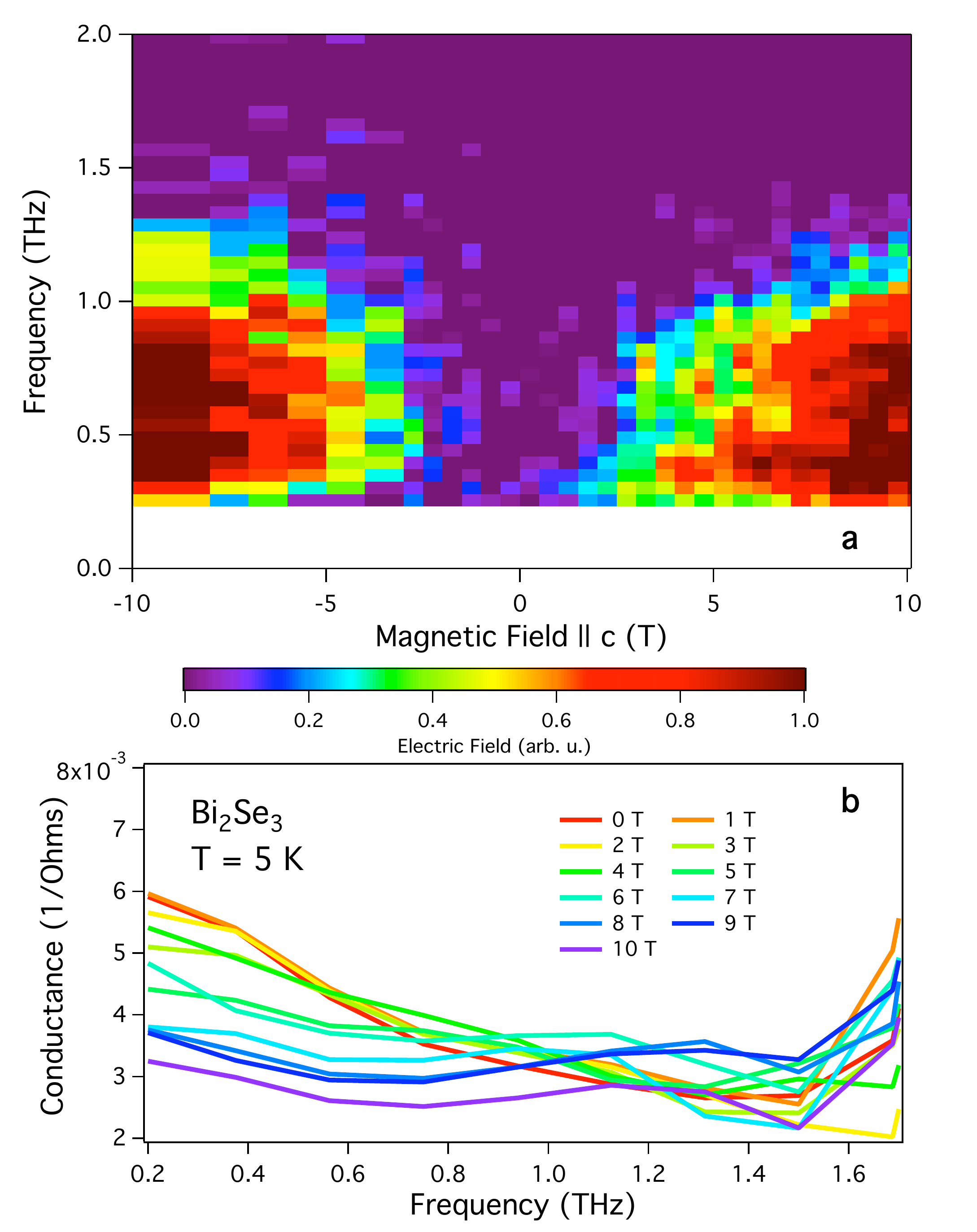}
\caption{\textbf{a}). Color map of the electric field magnitude of the 2$^{nd}$ THz pulse transmitted through the 16QL film in the cross polarizers configuration, for fields between -10 and 10 T applied along the film's \textit{c} axis (positive fields are antiparallel to the THz pulse propagation direction).
 \textbf{b}). Real part of the conductance of Bi$_2$Se$_3$ 100 QL film for magnetic fields between 0 and 10 T at 5 K.}
\label{Fig2}
\end{figure}

We now take advantage of a unique aspect of the time structure of TDTS and use the sapphire substrate itself as an optical resonator to measure both the Kerr and Faraday rotation angles in the same setup (further details in \cite{SuppMat}). In the experimental geometry of a thin film on a dielectric substrate, after transmitting through the sample, the THz pulse partially reflects back from the substrate-vacuum interface and returns to the film, where it reflects and travels back to the detector. In principle this process of internal reflection inside the sapphire substrate is repeated \textit{ad infinitum}. These \textit{echoes} are illustrated in Figs. \ref{Fig1}c and d, where we show typical time domain scans of a transmitted pulse and a schematic of the experimental configuration. In the presence of an external magnetic field $\mathbf{B}$, the first time the pulse is transmitted the wave polarization may be rotated an angle $\varphi_{F}$, the Faraday angle (labeled ``Faraday" in fig. \ref{Fig1}c).  As the 2nd pulse reflects back to the film and then reflects from it, the polarization may be rotated by an additional angle $\varphi_K$, the Kerr angle (labeled ``Kerr"). The fact that TDTS measurements are resolved in time allows the separation of the different contributions to the rotation angle; this type of separation is generally not possible with standard continuous wave techniques.

In Fig. \ref{Fig2}a we note the first evidence of an anomalously large Kerr rotation induced by the surface states in applied field. The figure shows the amplitude of the transmitted electric field from the Kerr pulse with crossed polarizers P1 and P3 at $\pm 45^\circ$ respectively (see \cite{SuppMat}). As we increase $\mathrm{B}$, we observe a rise in the transmitted amplitude as the electric field apparently undergoes a large rotation, since there should be no transmission  for the crossed polarizers unless the polarization is rotated by the sample. In contrast, the Faraday rotation was always small --at the level of the experimental sensitivity ($\sim$ 5$^\circ$)-- because it is proportional to the optically active layer thickness, which in this case is only a few nanometers. We neglect its contribution to the total Kerr $+$ Faraday rotation of the 2nd peak in the analysis below. We gain insight into the origin of this apparent large Kerr rotation by studying the longitudinal conductance G$_{xx}$ in an applied magnetic field $\mathbf{B}$. Within a conventional picture of cyclotron resonance, we expect that the spectral weight of the Drude term moves to higher frequency as the field is increased. The shift in spectral weight is also accompanied by the increase in the off-diagonal conductance G$_{xy}$ with magnetic field; this increase is responsible for inducing rotation to polarized light impinging on the sample.  One qualitatively observes such behavior in Fig. \ref{Fig2}b with a very weak maximum at finite frequency in the real part of the conductivity for the 100QL sample at 5 K and for several magnetic fields up to 10 T, and with the rotation indicated in Fig. \ref{Fig2}a. Films at other thicknesses show the same behavior.

\begin{figure}[h]
\includegraphics[width=.95\columnwidth]{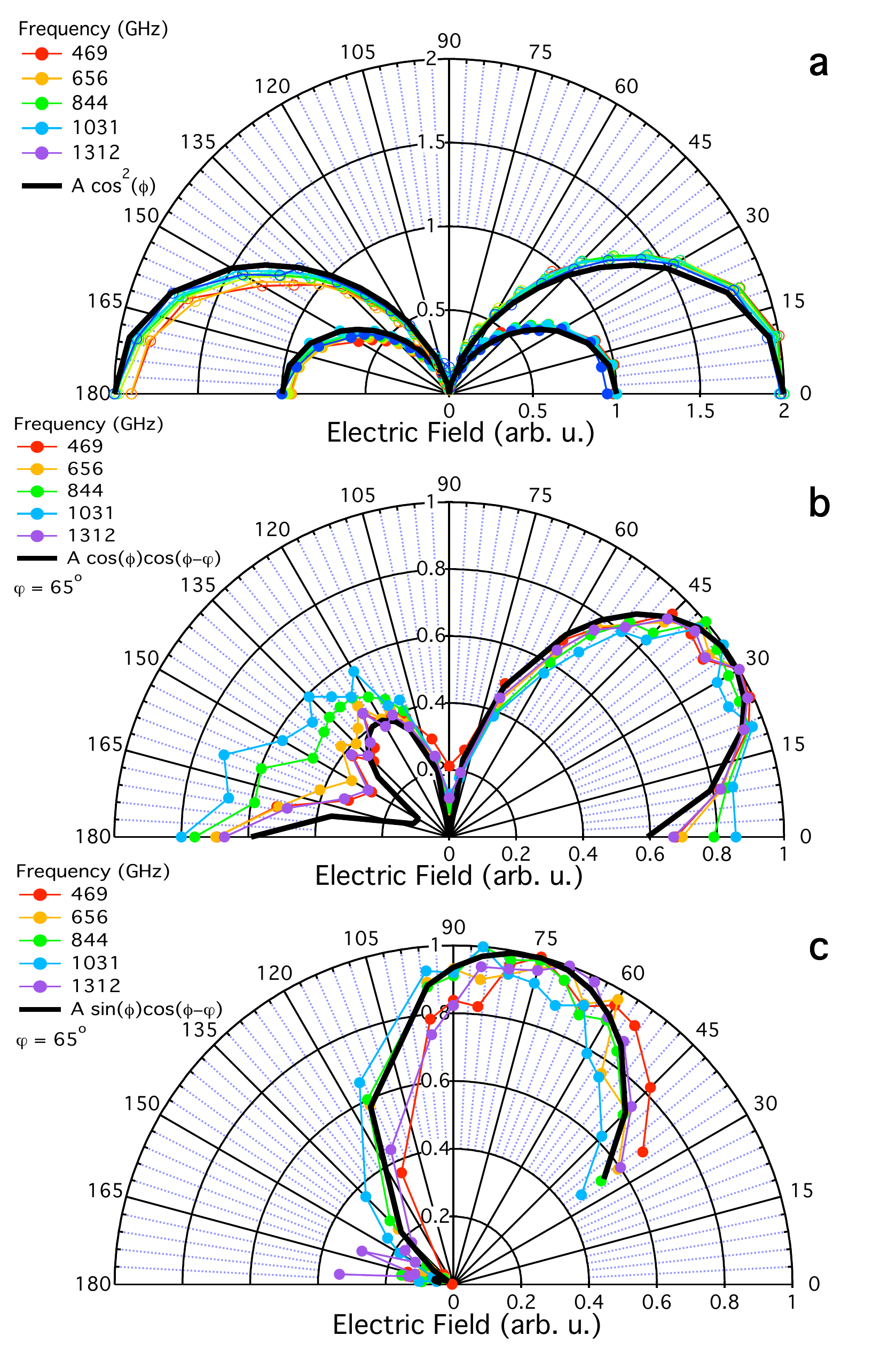}
\caption{Polarizer angle $\phi$ dependence of the transmitted electric field amplitude normalized to its maximum value at each displayed frequency for the 16 QL sample. All displayed data were taken at 10 T and 5 K. \textbf{a}). Amplitude of the transmitted electric field through a bare sapphire (Al$_2$O$_3$) substrate in the collinear polarizers configuration. Electric field of the Faraday pulse multiplied by 2 (open circles), and normalized amplitude of the Kerr pulse (closed) are shown. \textbf{b}). Normalized amplitude for the Kerr pulse in the collinear polarizer configuration. A rotation angle of 65$\pm 3^\circ$ is observed. \textbf{c}). Same as in \textbf{b}) in the cross polarizers configuration, also shows a rotation angle of 65$\pm 3^\circ$.}
\label{Fig3}
\end{figure}

We quantify the rotation angle in two different experimental configurations as described in further detail in ref. \cite{SuppMat}. In the \textit{collinear} polarizer configuration we expect the amplitude of the electric field to be proportional to $|\cos(\phi - \varphi)\cos(\phi)|$, and in the \textit{cross} polarizer mode $|\cos(\phi - \varphi)\sin(\phi)|$,  where $\phi$ is the angle of polarizer P2 and $ \varphi$ is the polarization induced by the sample. Fig. \ref{Fig3} shows precisely this behavior. Panel (a) shows a polar plot of the amplitude of the first and second transmitted pulses in the collinear arrangement through a bare substrate at 10 T and 5 K. It is clear that no rotation is observed and $\varphi$=0. In Figs. \ref{Fig3}b and c, we show polar plots for both collinear and cross polarizer geometries for a 16 QL sample also at 10 T and 5 K.  Both sets of data are consistent with a rotation angle of $\varphi_K$ = 65$\pm 3^\circ$. This is an extremely large value, both in its absolute scale and when normalized by the field and active thickness of the surface state. To the best of our knowledge, it is a world record for the Kerr rotation of a thin film.  

\begin{figure}[t]
\includegraphics[width=1\columnwidth]{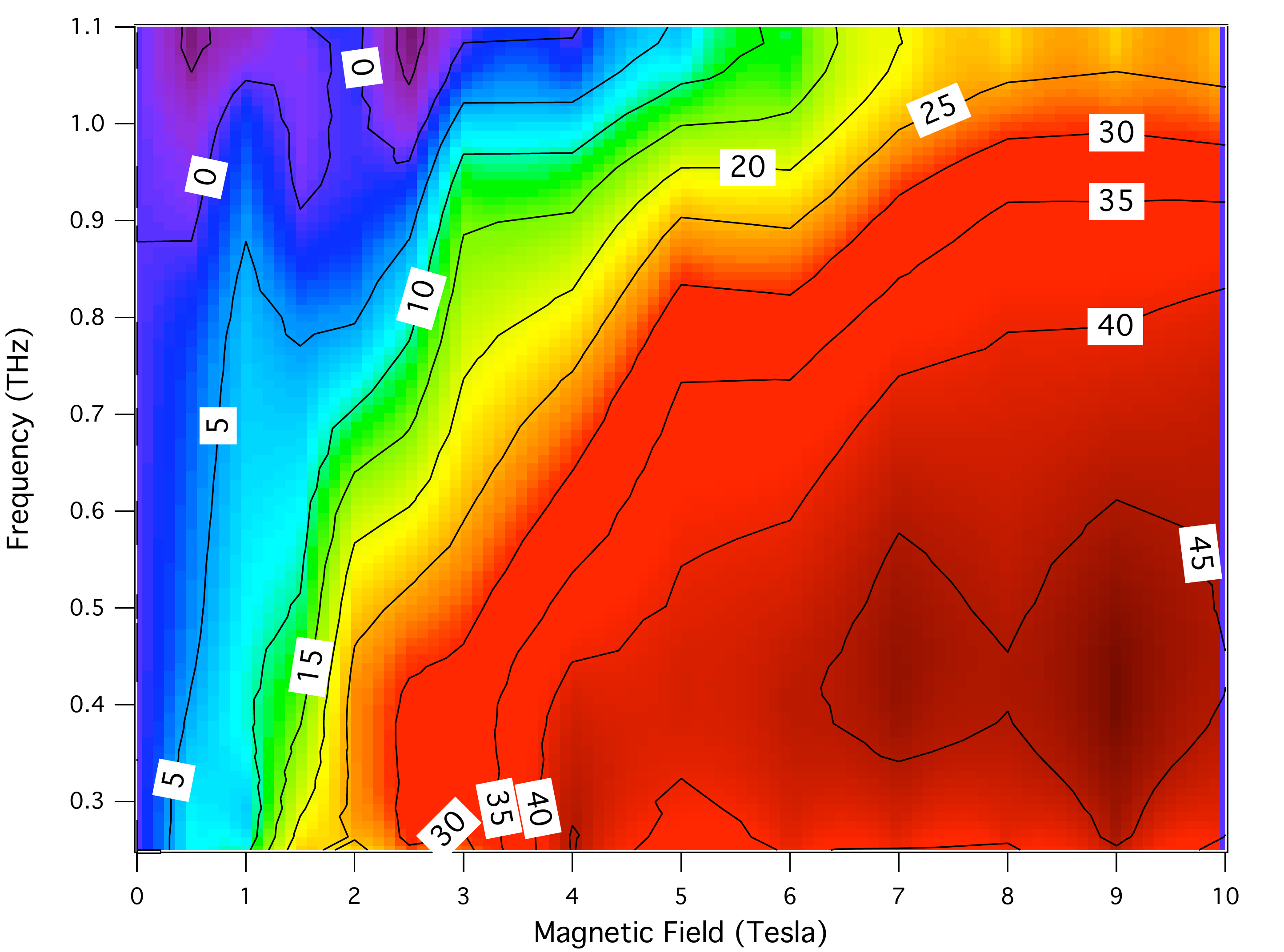}
\caption{Absolute value of Kerr rotation ($|\varphi_K|$) as a function of frequency and magnetic field for the 32 QL sample at 5 K. A maximum rotation of order 49.5$^\circ$ exists at low frequencies and finite fields.}
\label{Fig4}
\end{figure}

In Fig. \ref{Fig4}, we show the result of an experiment in a third configuration, where P2 is placed before the sample and is rotated at a high angular speed and the in- and out-of-phase outputs of a lock-in amplifier give the electric field components $\tilde X$ and $\tilde Y$ (more details in \cite{SuppMat}).  This method allows us to perform fast scanning of the magnetic field and frequency dependence of the Kerr angle.  In Fig. \ref{Fig4},  we show as a function of magnetic field and frequency the absolute value of the Kerr angle for the 32 QL sample measured at 5 K.  It is clear that a similar frequency dependence exists as in Fig. \ref{Fig2}b up to a maximum Kerr rotation of 49.5$^\circ$ for this sample. The variation in the values of the Kerr rotation with the filmÕs thickness, are consistent with the variations of the carrierÕs concentration and mobility as found in ref. \cite{Bansal11a}

Such a large Kerr rotation can be qualitatively explained using the physics of cyclotron resonance, if one correctly takes into account the enhancement that occurs when the reflection off the film happens from \textit{within} the substrate. The complex Kerr angle can be written as $\tan(\varphi_K)=\dfrac{2nZ_0\mathrm{G}_{xy}}{n^2-1-2Z_0\mathrm{G}_{xx}-Z_0^2(\mathrm{G}_{xx}^2+\mathrm{G}_{xy}^2)}$, where $n$ is the refractive index of the substrate, $Z_0 =$ 377 Ohms is the vacuum impedance and $\mathrm{G}_{xy}$ is the Hall conductance. We use parameters of the total conductance and scattering rate appropriate for our films  (see \cite{SuppMat} for details), and obtain from the fit an effective cyclotron mass, $m^*\sim$ 0.35 m$_e$. An estimate of the Dirac fermion cyclotron mass, $m^*=E_f/v_f^2$, using the Fermi energy from the carrier density n $\sim$ 3.3$\times$10$^{13}$cm$^{-2}$, $E_f \sim 0.5$eV, and velocity from ARPES \cite{Hsieh09b} $v_f\sim$5$\times$10$^{5}$ m/s, gives only a 10\% difference from our value. We find that this formula reproduces the main features of the data in Fig. \ref{Fig4}, as shown in Fig. S2 of ref. \cite{SuppMat}.  A similar analysis using bulk parameters \cite{Eto10a,Analytis10a,Butch10a} cannot reproduce the frequency and field dependence, which further confirms that the observed effect comes from the 2D surface states.  Small values of Kerr and Faraday rotations have been found recently on single crystals of Bi$_2$Se$_3$, which have an appreciable bulk contribution \cite{Jenkins10a}. Therefore, we can again conclude that the low frequency THz response is largely independent of the bulk contribution to the conductance and the observed colossal Kerr rotation is intrinsic to the surface metallic states.

It has recently been found \cite{Bianchi10,King11} that another type of 2D state exists in crystals of topological insulators cleaved in ultra-high vacuum. These states originate due to the band-bending effects, and as a result an electron accumulation layer emerges at the surface; its thickness was found to be $\sim$ 20 nm and the sheet carrier concentration of the order of 10$^{13}$ cm$^{-2}$. This implies that in thin films of less than 40 nm, transport would appear as effectively three dimensional. This is contrary to what we found in the THz conductivity, and to what is reported in DC transport where thickness-independent transport is found from 2 to 200 nm \cite{Bansal11a}. In addition, using the effective masses reported for these carriers ($m^* \sim$ 0.11 m$_e$ \cite{King11}), it is not possible to reproduce the frequency and field dependence of the Kerr angle as shown in ref. \cite{SuppMat}

We have shown clear evidence for the robust THz response of 2D topological surface states in thin films of topological insulator Bi$_2$Se$_3$. In magnetic fields we find a colossal Kerr rotation with angles up to $\varphi_K \approx$ 65$^\circ$ that is due to the cyclotron resonance of the surface electrons.  This measured Kerr rotation is larger by an order of magnitude than rotations typically found on high mobility GaAs heterostructures \cite{Jenkins10a}, but it is not quantized.  In order to reach a regime where the quantized topological magnetoelectric effect can be distinguished from semiclassical cyclotron resonance physics at accessible fields, it appears that the films will have to have an even larger mobility and the chemical potential must be tuned closer to the Dirac point. As the Fermi energy is tuned to the Dirac point, measurements like ours would have the clear signature of a reduction of the effective Dirac mass, and would eventually give way to the predicted signatures of \textit{axion} electrodynamics \cite{Qi08b}.  In general, the Kerr rotation we observe represents a benchmark for the intrinsic magnetoelectric effect predicted to exist in topological insulators. We also think that the ideas and techniques demonstrated in this work will be useful in studying effects of the interaction of polarized THz radiation with other novel states of matter.

$Acknowledgements.$
The authors would like to thank H.D. Drew, J. Hancock, Z. Hao, G.S. Jenkins, A. Kuzmenko, A. MacDonald, N.A. Mecholsky, A.J. Pearson, O. Tchernyshyov, W-K. Tse, and Y. Wan for helpful discussions and/or correspondences.   Support for the measurements at JHU was provided under the auspices of the ``Institute for Quantum Matter" DOE DE-FG02-08ER46544 and the Gordon and Betty Moore Foundation. The work at UB was supported by NSF MRI-R2 53383-1-1085743 and NSF DMR-1006078.   The work at Rutgers was supported by IAMDN of Rutgers University, NSF DMR-0845464 and ONR N000140910749.


\renewcommand{\thefigure}{\Alph{figure}\arabic{addendumfig}}
\setcounter{figure}{18}
\setcounter{addendumfig}{1}

\section {Supplemental Material}
\subsection {Film growth.}

Thin films of topological insulator Bi$_2$Se$_3$ of varied thicknesses were grown at Rutgers by a molecular beam epitaxy technique on 0.5 mm thick sapphire substrates (Al$_2$O$_3$).  Films were grown on ozone-cleaned surfaces using the two-temperature growth process. Evolution of the film surface during growth was monitored by RHEED.  After deposition of 3 QL of Bi$_2$Se$_3$ at 110$^\circ$ C, a sharp streaky pattern was observed, indicating the growth of single-crystal Bi$_2$Se$_3$ structure. The film was then slowly annealed to a temperature of 220$^\circ$ C, which helped further crystallization of the film as seen by the brightening of the specular spot. The diffraction pattern and the Kikuchi lines became increasingly sharp on further deposition. This shows that the grown films have atomically flat morphology and high crystallinity.  This process led to high quality single crystalline films with the largest terraces, highest bulk mobilities, and lowest volume carrier densities as detailed elsewhere \cite{Bansal11a}.

\subsection{THz Methods}

TDTS measurements in zero magnetic field were performed at JHU using a home-built transmission based time-domain THz spectrometer. In this technique, an infrared femtosecond laser pulse is split into two paths and sequentially excite a pair of photoconductive `Auston'-switch antennae on radiation damaged silicon on sapphire. A broadband THz range pulse is emitted by one antenna, transmitted through the Bi$_2$Se$_3$ film, and measured at the other antenna.  By varying the length-difference of the two paths, the electric field of the transmitted pulse is measured as a function of time.  Ratioing the Fourier transform of the transmission through the Bi$_2$Se$_3$ film on a substrate to that of a bare reference substrate we resolve the frequency dependent complex transmission of the film.  The transmission is inverted to obtain the complex conductance by the standard formula for thin films on a substrate: $\tilde{T}(\omega)=[(1+n)/(1+n+Z_0\tilde{\mathrm{G}}(\omega))] e^{i\Phi_s}$ where $\Phi_s$ is the phase accumulated from the small difference in thickness between the sample and reference substrates, $n$ is the substrate index of refraction, $Z_0 \approx$ 377 $\Omega$ is the vacuum impedance, and $\tilde{\mathrm{G}}=\mathrm{G}^{'}_{xx}+i\mathrm{G}^{''}_{xx}$ is the film's effective complex conductance. In the case of topological insulators, the effective conductance is composed of a sum of a bulk contribution, $\tilde{\mathrm{G}}_{bulk}$=${\sigma}_{bulk}\times t$ where $t$ is the film thickness, and a surface contribution 2$\tilde{\mathrm{G}}_{surface}$.

Experiments in magnetic field were done in a similar fashion via TDTS spectroscopy at UB. The detection of THz radiation in this system is achieved by electro-optic means. In this method a ZnTe crystal is impinged upon by a part of the infrared femtosecond laser; when the THz pulse reaches the ZnTe crystal, it becomes birefringent and changes the polarization characteristics of the infrared beam. The changes in the polarization of the infrared ultrafast laser beam, which are proportional to the THz electric field strength, are measured by balanced photodiode detection. 

As shown in Fig 1c, measurements in field were done in two configurations using three wire grid polarizers of THz radiation (P1, P2 and P3). P1 was placed before the sample to ensure linearly polarized light was incident to the film, P2 and P3 were placed after the sample where P2 was placed in a standard rotating holder that allowed the selection of the polarizer angle ($\phi$). In the first setup we measured the transmitted amplitude as a function of P2's angle  $\phi$, with polarizers P1 and P3 parallel to each other at 0 degrees (collinear configuration).  In this mode, we expect the amplitude of the electric field to have the dependence on $\phi$ as $|\cos(\phi - \varphi)\cos(\phi)|$, where $\varphi$ is the pulse rotation angle measured with respect to the position of the first polarizer.  In the second configuration, P1 and P3 are perpendicular to each other, at 45$^\circ$ and -45$^\circ$ from the vertical (cross polarizer configuration), and again we measure the intensity as a function of $\phi$. In this case the amplitude of the electric field should have an angle dependence as  $|\cos(\phi-\varphi)\sin(\phi)|$. In the cross polarizer configuration, the first polarizer had an angle of 45$^\circ$ with respect to the collinear configuration due to the sensitivity of the electro-optic detection to the polarization of the pulse. 

The third experimental configuration used for complete characterization of the Kerr angle in the frequency-magnetic field plane was done using a rotating polarizer technique. P2 was held in a fast rotating stage and was placed between P1 and the sample, with polarizers P1 and P3 both collinear and oriented vertically (0$^\circ$). The fast rotator is spun at approximately 600 rpm, and the resulting signal is demodulated by a lockin amplifier at twice the frequency set by the rotation speed.   In such an experiment the in-phase signal of the lock-in is proportional to the transmitted electric field collinear with the incoming polarization ($X$ signal), and the out-of-phase response is proportional to the electric field at 90$^\circ$ from the original polarization ($Y$ signal) \cite{Grayson02a,Markowicz04}.  This method allows fast magnetic field scanning and the analysis of the full spectral response.  The $X$ and $Y$ time dependent signals are Fourier transformed to obtain $\tilde X(\omega)$ and $\tilde Y(\omega)$. By taking the ratio of  $\tilde Y(\omega)$ to $\tilde X(\omega)$, we obtain the tangent of the rotation angle as $\tan[\varphi(\omega)]$=$\frac{\tilde Y(\omega)}{\tilde X(\omega)}$.  In this work we plot the absolute value of $\varphi$.

\subsection {Temperature dependent conductance.}

\begin{table*}[h t]
\caption{Fitting parameters for the Drude and phonon contributions to the conductance at zero field and 6 K.}
\begin{ruledtabular}
\begin{tabular}{lccccccc}
  & $(\omega_{pD}/2\pi)^2\times$ t & $\gamma_D= \Gamma_D/2\pi$ & $(\omega_{pP}/2\pi)^2\times$ t & $\omega_{P}/2\pi$ & $\gamma_P=\Gamma_P/2\pi$  & $\varepsilon_{\infty}\times$ t\\
  & (THz$^2\times$nm)      & (THz)   & (THz$^2\times$nm)    & (THz)            & (THz) & (nm) \\
\hline
 16QL & 193,382        & 1.41 &  4,067    & 1.89                   & 0.108                 & 38,592  \\
 32QL   & 201,014       & 1.37 &  14,440    & 1.92                 & 0.094                 & 103,793 \\
 64QL    &  148,818      & 1.01 & 35,867   & 1.89                    & 0.081                & 19,664 \\
 100QL & 157,253       & 1.02 &   53,124   & 1.90                  & 0.086               & 61,975 \\
\end{tabular}
\end{ruledtabular}
\end{table*}

We fit the zero field THz conductance data using a 2D model dielectric constant consisting of two identical Drude terms for the free electron response, a Drude-Lorentz oscillator for the phonon, and a dielectric constant ($\varepsilon_\infty$) that represents the high frequency optical transitions, $\varepsilon_{xx}=1+\frac{4\pi i\sigma_{xx}}{\omega}$, where $\sigma_{xx} = \mathrm{G}_{xx} / t$ with $t$ the film thickness:
\begin{equation}
\centering
\tag{S1}
\varepsilon_{xx}=\varepsilon_{\infty}-\frac{\omega_{pD}^2}{\omega^2-i\omega\Gamma_D}+\frac{\omega_{pP}^2}{\omega_P^2-\omega^2-i\omega\Gamma_P}
\label{Drude}
\end{equation}

\begin{figure*}[t]
\includegraphics[width=1.5\columnwidth]{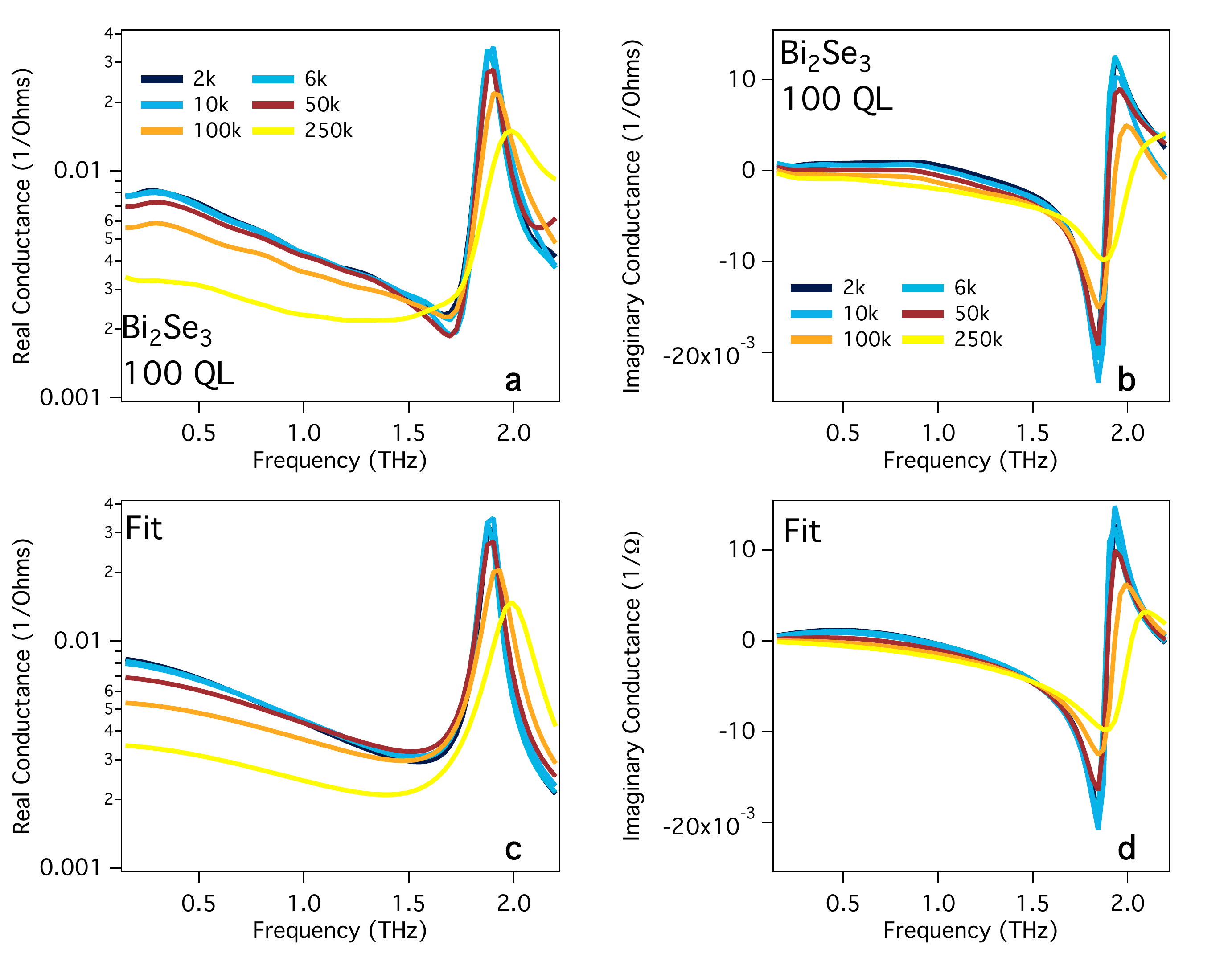}
\caption{Frequency and temperature dependent conductance. \textbf{a}). Measured real and \textbf{b}) imaginary conductances for the 16 QL sample for the displayed temperatures. \textbf{c}) and \textbf{d}). Respective fits using the model described in equation \ref{Drude}.}
\label{Fig:S1}
\end{figure*}

\noindent Here the subscripts $D$ and $P$ represent the Drude and the phonon contributions, respectively. We measured 3 samples of thicknesses 16, 32, 64 and 100 QL (1 QL $\sim$ 0.94 nm) in the temperature range between 2 and 300 K. Fig. \ref{Fig:S1} shows the real and imaginary conductances at several temperatures for the 100QL sample. The fit parameters are given in Table 1.

\subsection{Calculation of Kerr angle for a thin metallic film.}
The Faraday ($\phi_F$) and Kerr ($\phi_K$) angles are defined as:
\begin{eqnarray*}
\tan(\varphi_F) = i\frac{t_+-t_-}{t_++t_-}\\
\tan(\varphi_K) = i\frac{r_+-r_-}{r_++r_-}
\end{eqnarray*}
\noindent where $t_\pm$ ($r_\pm$) is the transmission (reflection) coefficient for (+) right circularly polarized and (-) left circularly polarized light.
In the case of the wavelength of light being much larger than the thickness of the film, we obtain the reflection and transmission coefficients as follows.
\begin{eqnarray*}
r_\pm = \frac{n-1-Z_0\mathrm{G}_\pm}{n+1+Z_0\mathrm{G}_\pm}\\
t_\pm = \frac{2}{n+1+Z_0\mathrm{G}_\pm}
\end{eqnarray*}
\noindent where $\mathrm{G}_\pm = \mathrm{G}_{xx} \pm i\mathrm{G}_{xy}$ are the (+) right and (-) left circularly polarized conductances. The conductances have the usual form for the free carrier response in a magnetic field as follows:

\begin{eqnarray*}
\mathrm{G}_{xx}=\frac{\omega_{pD}^2/(4\pi)(\Gamma-i\omega)}{(\Gamma-i\omega)^2-\omega_{c}^2}\\
\mathrm{G}_{xy}=-\frac{\omega_{pD}^2/(4\pi)\omega_{c}}{(\Gamma-i\omega)^2-\omega_{c}^2}
\end{eqnarray*}
where $\omega_c = e\mathbf{B}/m^*c$ is the cyclotron frequency.

We can then simplify for the Kerr angle when reflection happens \textit{within} the substrate as:
\begin{equation*}
\tag{S2}
\tan(\varphi_K)=\frac{2nZ_0\mathrm{G}_{xy}}{n^2-1-2Z_0\mathrm{G}_{xx}-Z_0^2(\mathrm{G}_{xx}^2+\mathrm{G}_{xy}^2)}
\label{Kerr}
\end{equation*}

Similarly, the Kerr angle can be calculated for the case of reflection from the vacuum-TI interface:
\begin{equation*}
\tag{S3}
\tan(\varphi_K)'=\frac{2Z_0\mathrm{G}_{xy}}{1-n^2-2nZ_0\mathrm{G}_{xx}-Z_0^2(\mathrm{G}_{xx}^2+\mathrm{G}_{xy}^2)}
\label{Kerr2}
\end{equation*}

We note that among the differences between equations \ref{Kerr} and \ref{Kerr2} are the appearance of the substrate's index of refraction $n$ in the numerator, which makes the rotation larger by this factor, as well as its multiplication in the denominator of the factor $-2nZ_0\mathrm{G}_{xx}$ in eqn. \ref{Kerr2} which makes its rotation smaller (by making the denominator larger). In the quantum Hall effect regime that is relevant for the topological magnetoelectric effect (G$_{xx} \sim0$ and G$_{xy} \sim e^2/h$) these expressions can be simplified to:

\begin{eqnarray*}
\tan(\varphi_K)\sim\frac{2nZ_0\mathrm{G}_{xy}}{n^2-1} = \frac{n\alpha}{n^2-1}\\
\tan(\varphi_K)'\sim\frac{2Z_0\mathrm{G}_{xy}}{1-n^2} = \frac{\alpha}{1-n^2}
\end{eqnarray*}
where $\alpha$ is the vacuum fine structure constant. Thus, it is clear that making the reflection measurement from \textit{within} the substrate leads to an enhancement of the Kerr angle by the substrate's index of refraction.

\setcounter{figure}{18}
\setcounter{addendumfig}{2}

\begin{figure*}[t]
\begin{center}
\includegraphics[width=1.9\columnwidth]{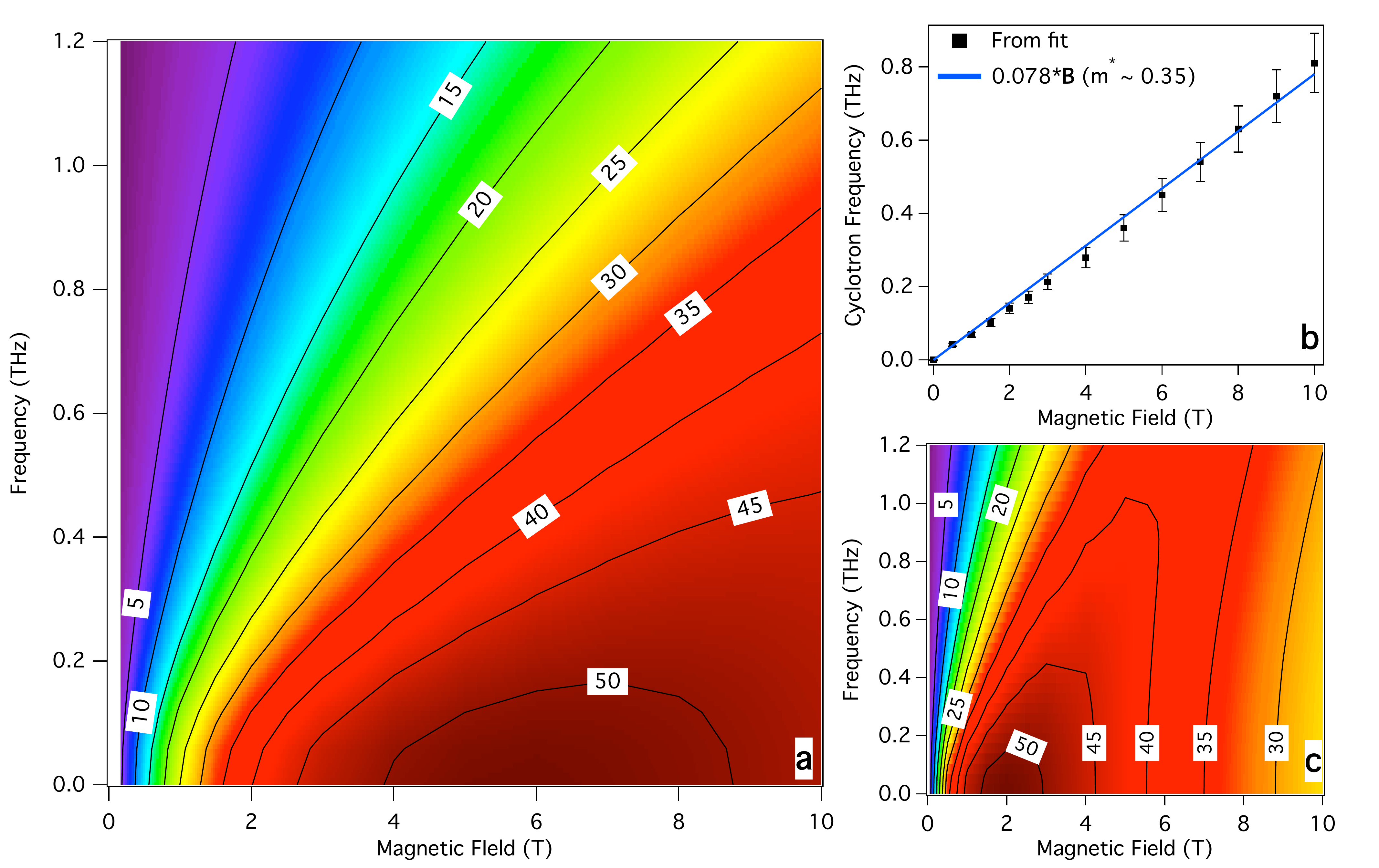}
\end{center}
\caption{Calculation of Kerr angle and cyclotron frequency. \textbf{a)} Calculated absolute value of the Kerr angle ($|\varphi_K|$) as a function of frequency and magnetic field using the parameters given in the text. \textbf{b)} Cyclotron frequency obtained from fitting the low frequency part of the spectrum (symbols), line is obtained by taking the effective mass $m^*\sim$ 0.35 m$_e$. \textbf{c} Calculation of the Kerr angle for an effective mass of $m^*\sim$ 0.12 m$_e$.}
\label{Fig:S2}
\end{figure*}

With equation \ref{Kerr} we can reproduce the magnitude, frequency and field dependence of the  measured Kerr angle with an almost linear dependence of the cyclotron frequency on magnetic field, as shown in Fig. \ref{Fig:S2}(a), with an effective mass $m^*\sim$ 0.35 m$_e$ obtained from the cyclotron frequency in Fig. \ref{Fig:S2}(b). As discussed in the main text, taking an effective mass of 0.12 m$_e$, which is the one obtained for the states induced by band-bending near the surface \cite{King11}, one cannot reproduce the results of the experiment as shown in Fig. \ref{Fig:S2}(c). Similarly poor comparison with the data can be obtained when the bulk effective mass, $m^*\sim$ 0.16 m$_e$, is used.

\bibliography{TopoIns}


\end{document}